# Raman spectroscopy of sodium chloride under high-pressure and high-temperature

Yu Tian[a,b,c], Wansheng Xiao[a,b,]*, Yunhong He[a,b,c], Huifang Zhao[a,b,c], Feng Jiang[a,b,c], Dayong Tan[a,b], Ming Chen[a,b]

*a CAS Key Laboratory of Mineralogy and Metallogeny, Guangzhou Institute of Geochemistry, Chinese Academy of Sciences, 511 Kehua Street, Guangzhou 510640, China;*

*b Key Lab of Guangdong Province for Mineral Physics and Materials, 511 Kehua Street, Guangzhou 510640, China;*

*c University of Chinese Academy of Sciences, 19 Yuquan Road, Beijing 100049, China*

*Corresponding author.

*E-mail addresses:* wsxiao@gig.ac.cn (W. Xiao)

## ABSTRACT

The high-pressure and high-temperature behaviors of sodium chloride (NaCl) have the fundamental and application significance to the high pressure physics and chemistry. To explore the reactivity of NaCl at high pressure and high temperature, we compressed the NaCl samples to pressures up to 85 GPa and heated them to a temperature of 1800±300 K by using the diamond anvil cell and the double-sided laser heating technology. Our investigation of Raman spectroscopy shows that NaCl decomposes to $NaCl_3$ with orthorhombic *Pnma* structure plus possible $Na_3Cl$ with tetragonal *P*4/*mmm* structure above 31 GPa. It shows that the competition of the unusual Cl-Cl covalent bond and Na-Na metallic bond with the Na-Cl ionic bond at high pressure and high temperature may prefer the occurrence of the $NaCl_3$ and $Na_3Cl$ combination. We speculate that the large temperature gradient in the NaCl sample layers in the diamond anvil cell experiments, which indicate a non-equilibrium state, is critical to the decomposition reaction of NaCl observed in this study. The catalytic effect of the used laser absorbing materials of transition metal oxides could be another factor to promote the chemical reaction of NaCl. The results of this investigation provide a new pathway to investigate the distinctive non-1:1 stoichiometric alkali metal-halogen compounds that may have exotic properties.

1. Introduction

Sodium chloride, NaCl, a prototype of ionic crystal, has been investigated extensively from various experimental and theoretical aspects [1-12]. The conventional ionic bond model indicates that the nature of Na (Cl) atom prefers to lose (gain) an electron to form monovalent cation (anion) with stable noble gas electron configuration agreeing with the octet rule. The primary ionicity of Na-Cl bond dominates the structure and properties. The charge neutrality rule constrains the Na-Cl system to form only one stable bulk solid of NaCl with 1:1 stoichiometry at ambient conditions. The electrostatic interaction between the closed-shell $Na^+$ cation and $Cl^-$ anion leads to the formation of a crystal in which the ions are stacked in an alternating fashion along the three dimensions to form the cubic structure (B1, $Fm3m$). The high lattice energy (786 kJ mol$^{-1}$) [13] demonstrates the prominent stability of NaCl crystal.

Recently, the experimental and theoretical studies indicate the stability of various specimens of Na-Cl system at unconventional conditions [14-16]. The 2D crystals of $Na_2Cl$ and $Na_3Cl$ were observed depositing at the surface of graphene and graphite from the NaCl dilute solution [14]. A series of Na-Cl compounds of different stoichiometries were predicted to be stable at high pressures, and $NaCl_3$ and $Na_3Cl$ bulk solids have been synthesized by chemical reactions of NaCl + $Cl_2$ and NaCl + Na at high-pressure and high-temperature (HPHT) conditions [15]. These compounds display the various atomic interactions of Na-Na metal bond, Cl-Cl covalent bond as well as Na-Cl ionic bond. The analogous compounds of alkali halides with unusual stoichiometry such as $KCl_3$, $K_3Cl$ and $KBr_3$ have also been synthesized under pressure

[17, 18]. The atomic arrangement of these non-1:1 stoichiometric compounds resembles to the geometry of the low-dimensional specimens of alkali halides such as $Na_mCl_n$ clusters and $NaCl_3$ molecule [19-23]. It is worth highlighting that the nonlinear and unsymmetrical trichloride anion ($Cl_3^-$) in the $NaCl_3$ bulk solid is consistent with that of in the $NaCl_3$ molecule [23-25]. It seems that the competition of various chemical bonds in the Na-Cl system is responsible for the stability of non-1:1 stoichiometric compounds at unconventional conditions.

The high pressure behaviors of NaCl have been studied generally due to the fundamental significance as well as the wide technological applications such as the pressure scale [26, 27] and thermal insulator [28-30] in the HPHT diamond anvil cell (DAC) experiments. It transforms to the CsCl-type structure (B2, $Pm3m$) at about 30 GPa and remains stable up to 300 GPa at room temperature [31, 32]. At higher pressure above 300 GPa, the NaCl B2 phase was predicted changing to the orthorhombic polymorphs of oC8, oI8 and oP16 structures, in which the identical ions of both Na and Cl bond to form the chains, layers or frameworks and display the metal conductivity [33]. Although the chemical properties and bonding behaviors of Na and Cl may alter drastically at extreme pressure, it is generally considered that NaCl remains stable in the wide pressure range and will not spontaneously decompose into other compounds [15]. As a result, the existing synthesized experiments of different stoichiometric alkali halides employed the reactions of stoichiometric alkali halides with excess of either alkali metal or halogen [15, 17, 18].

The previous HPHT experimental studies on NaCl focused mainly on the pressure (P) - melting temperature (T) relationship [6, 34-36], the B1-B2 P-T phase boundary [37, 38], and the P-V-T equation of state [39, 40]. The identification of specimens by X-ray diffraction (XRD) in these experiments was restricted to either pressure lower than 30 GPa or temperature no more than 1000 K. Raman spectroscopy was absent in these kinds of study. It is noticed that, however, there were actually tens or hundreds of laser-heating HPHT experiments using NaCl as thermal insulator and pressure scale to carry out the XRD measurement at wide pressure and temperature ranges, for example, Shen et al [41]. Almost all of these studies indicated the stability of NaCl in equilibrium. Nevertheless, there were a few unusual observations. Dewaele et al [29] reported the XRD evidences of Ta reacting with NaCl at HPHT conditions. The reaction products remain unknown. Recently, we conducted the experimental study on chemical reaction of NaCl with $O_2$ at high pressure of about 55 GPa and high temperature of 1500-2000 K. The Raman measurements validated the unanticipated occurrence of $NaCl_3$ [42]. It is obviously different from the recognition that alkali metal perchlorates, such as $NaClO_4$ and $KClO_4$, which are usually used as oxidizing agents in material synthesized experiments [43,44], decompose to alkali halides and $O_2$ at HPHT conditions [45]. Those unusual observations prompt us to explore the stability and chemical reactions of NaCl at certain HPHT environments. Here we report the Raman observations revealing clearly the chemical reaction of NaCl at HPHT.

## 2. Experimental details

The NaCl powder with 99.99% purity was dried in an oven at ~400 K for more than 24 hours before used as experimental sample. It displays no Raman band in accord with group theory analysis of its B1 phase. The HPHT experiments were conducted by using the symmetric Mao-Bell-type DAC and the micro-laser-heating equipment. The diamond anvils with 300-μm and 200-μm culet were used to generate pressure below 60 GPa and 85 GPa, respectively. T301 stainless steel gaskets were indented to a thickness of 30(25) μm and drilled a 95(65) μm diameter hole for the 300(200)-μm diamond-anvil experiments. A piece of laser-absorbing material (LAM) about one third of the sample hole diameter was sandwiched between NaCl sample layers and loaded into the sample chamber. A few fine ruby chips were loaded as pressure marker [46]. The samples were pressed directly to the desired pressures at room temperature, and then heated from both sides by a 100 W SPI fiber laser of 1.07-μm wavelength. The laser beam was focused to a spot of ~20 μm and scanned on the LAM to heat the NaCl sample for about 10 min. Thermal radiation spectra from both sides of the sample were measured using a CCD imaging spectrometer and fitted to the Planck black-body radiation function to determine the temperatures [28]. The experimental temperatures were measured to be 1800±300 K. We permitted the obvious temperature uncertainty since the sample configuration described above inevitably results in obvious temperature gradient in the NaCl sample [47-50].

It is necessary to use the LAM to heat the NaCl sample since NaCl is transparent to the near-infrared laser beam. We have used $Fe_2O_3$, $Cr_2O_3$, FeO, $MgMnO_3$, and Pt as LAMs in this study. The identical Raman spectra assured the measured data were independent of the LAMs. In the following, the displayed Raman data were obtained from the $Fe_2O_3$ as absorbing material experiments except the special statement.

Raman spectra were measured after the experimental samples were heated and quenched to room temperature. A Renishaw 2000 micro-Raman spectrometer with exciting lasers of 532, 633, and 785 nm wavelength was used to record the spectrum in the backscattering geometry. Unless otherwise specified, the 532 nm laser was used as exciting light source in this study. A 20× long working distance objective was used to focus the exciting laser beam to ~3 μm diameter and to collect the Raman scattering. A thermoelectrically cooled CCD detector was equipped to collect the scattered light dispersed by a 1800 lines/mm grating with a resolution of ~1 $cm^{-1}$. The measured spectrum range is 100-1200 $cm^{-1}$ and the acquisition time is 10 s for each spectrum.

## 3. Results and discussions

### 3.1. Raman features of the thermal-treated NaCl at high pressures

Fig. 1 show three spectra measured from the same sample with different exciting light sources of 532, 633, and 785 nm wavelength at a pressure of 54.4 GPa and room temperature. The initial NaCl solid as the sample has been heated at the pressure. All of the three spectra present a series of strong bands. It can be seen clearly that the three spectra exhibit completely consistent characteristics. It manifests the observed spectra are independent of the wavelength of the excitation light used. The fact means that the measured spectra are the Raman scattering reflecting the lattice vibrations of the sample rather than other possible sample luminescence. The Raman spectrum of the sample at this pressure shows nine vibration bands, and the frequencies are shown in Table 1.

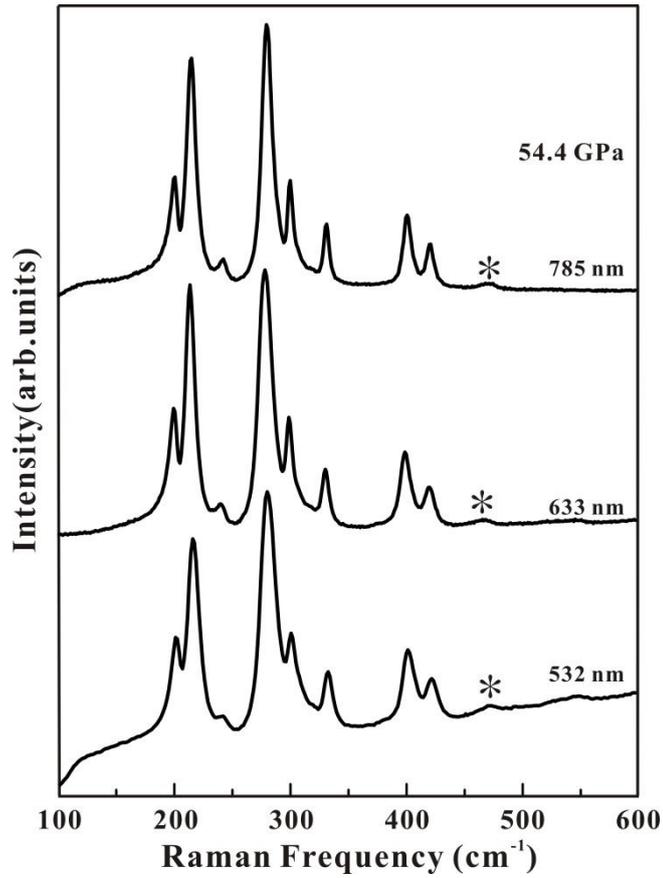

**Fig. 1.** Raman spectra of the annealed NaCl sample measured at 54.4 GPa and room temperature with different exciting light sources of 532, 633, and 785 nm wavelengths respectively. The complete consistence of these spectra indicates the chemical change of NaCl at HPHT. About the bands marked by asterisk symbols, please see text.

**Table 1** Raman frequencies of the experimental spectra of *Pnma*-NaCl$_3$.

|  | 54.4 GPa | 57.2 GPa | 57 GPa [15] |
|---|---|---|---|
|  | 202 | 204 | 204 |
|  | 216 | 220 | 219 |
|  | 241 | 244 | 243 |
|  | 281 | 287 | 286 |
| NaCl$_3$ | 302 | 307 | 305 |
|  | 333 | 338 | 338 |
|  |  | 378 |  |
|  | 403 | 412 | 409 |
|  | 423 | 432 | 428 |
| Na$_3$Cl | 474 | 488 |  |

NaCl solid transforms from B1 phase to B2 phase (*Pm3m*) at pressure of ~30 GPa [31] and remains stable up to at least 300 GPa at room temperature [32]. Group theory predicts that the B2 phase of NaCl has no Raman-active vibration and only one infrared-active band ($T_{1u}$). It agrees with the experimental observation when we compressed the NaCl sample to 54.4 GPa before heating. The observed Raman spectra of the sample shown in Fig. 1 are obviously not resulted from the vibration of the NaCl B2 phase, and therefore, there must be other Na-Cl compounds in the sample system responsible for the measured Raman scattering. It is clear that these Na-Cl compounds are produced by the chemical reaction of NaCl at the experimental HPHT environments. This observation is definitely different from most of the reported experimental measurements about the stability of NaCl at HPHT.

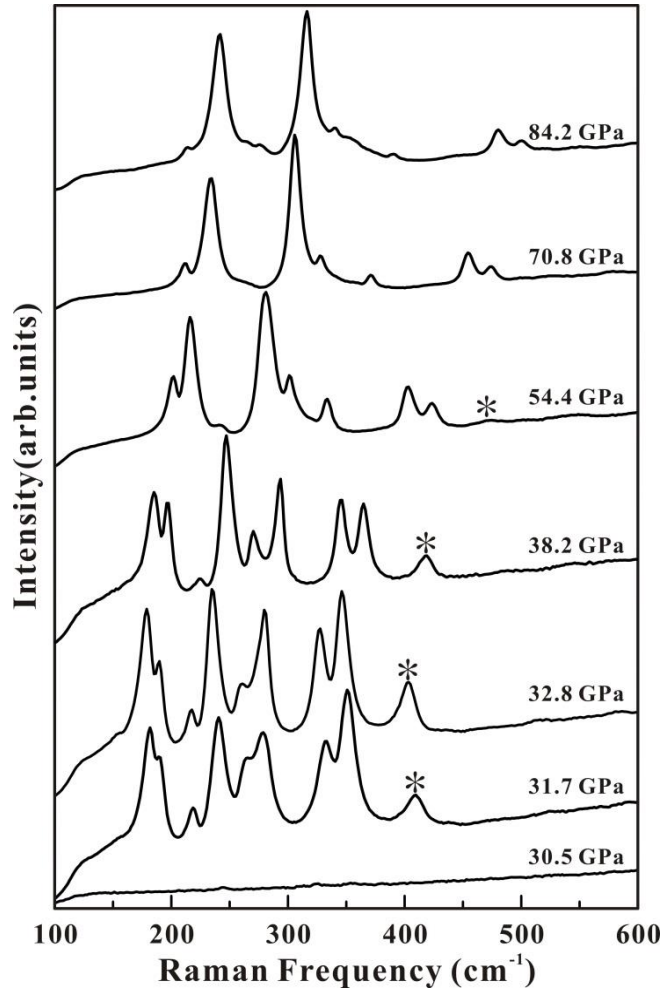

**Fig. 2.** Raman spectra of the NaCl sample observed at various pressures from 30.5 GPa to 84.2 GPa after laser-heating. It suggests the chemical reaction of NaCl occurs over a wide pressure range above 31 GPa at high temperature. About the bands marked by asterisk symbols, please see

text.

Fig. 2 displays the Raman spectra measured at room temperature after NaCl was heated at various pressures. The 30.5 GPa spectrum presents almost a flat line in which the very weak Raman bands are originated from the LAM of $Fe_2O_3$ hematite. It manifests the NaCl B2 Phase remain stable after heating in the pressure consistent with the traditional understanding. However, when the sample was compressed to 31.7 GPa and then heated, a series of strong Raman bands emerged. It indicates that the NaCl sample undergoes chemical change at this pressure. We further heated the sample at 32.8 GPa and 38.2 GPa respectively, the Raman measurements obtained nearly the same as the 31.7 GPa spectrum reflecting the steady occurrence of the Na-Cl compounds at pressures higher than ~31 GPa after heated. We also performed the various experiments directly squeezing the samples to the highest experimental pressures of 54.4 GPa, 70.8 GPa and 84.2 GPa respectively. After heating, the measured Raman spectra, as shown in Fig. 2, exhibit distinct Raman bands obviously different from the flat line spectra obtained before heating. These experimental observations reveal definitely that the chemical reaction of NaCl can occur over a wide pressure range above 31 GPa at high temperature in our experiments.

### *3.2. Products of NaCl chemical reaction at high pressure and high temperature*

We noticed that the Raman spectrum consistent with measured in this study was presented in two literatures [15, 30]. Fig. 3 shows the Raman spectra of 54.4 GPa (curve a) and 57.2 GPa (curve b) measured in this study in comparison with the 57 GPa spectrum (curve c) shown in supplementary materials Fig. S9 in reference 15, which was observed in the NaCl + $Cl_2$ experiment and ascribed to the *Pnma*-$NaCl_3$ confirmed by XRD measurement. The band frequencies of the three experimental spectra shown in Fig. 3 were presented in Table 1. The nearly identical spectral profiles of curve b and curve c suggest that the experimental products of NaCl chemical reaction at HPHT in this study contain the *Pnma*-$NaCl_3$.

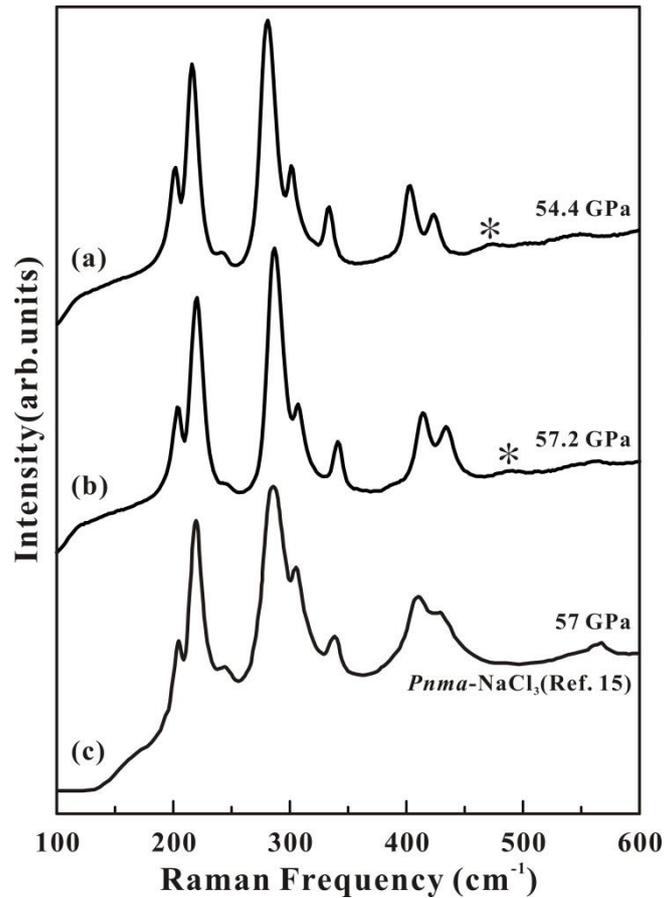

**Fig. 3.** Experimental Raman spectra of the *Pnma*-NaCl$_3$ at similar pressure for comparison. Experimental spectra (a) at 54.4 GPa and (b) at 57.2 GPa were collected in this work. Experimental spectrum (c) at 57 GPa is cited from reference 15. About the bands marked by asterisk symbols, please see text.

Fig. 4 presents several representative Raman spectra measured at various pressures on decompression after annealing at about 85 GPa. It can be seen that the relative intensity of the bands changes obviously at different pressures. Nevertheless, all the Raman bands can be traced from 85 GPa to below 15 GPa. When the pressure is decreased to lower than about 20 GPa, the Raman bands of *Pnma*-NaCl$_3$ change dramatically by losing the intensity and increasing the width indicating its instability. It agrees with the decomposition reaction of NaCl$_3$ observed previously [15, 42].

In addition, it can be seen that a band marked by asterisk in several measured Raman spectra in Fig. 4 exhibits different behavior comparing with the other bands as pressure changes. This particular band is also shown in some other Figures at various pressures. It is merely visible at pressure lower than 70 GPa and remains weak to about 40 GPa. It then increases the intensity as pressure decreases to about 15 GPa, and become weak again and disappear finally at about 9 GPa. The pressure behavior of this band is clearly distinct from the other bands suggesting it is neither the fundamental vibration nor the overtone of the *Pnma*-NaCl$_3$. We consider that it is attributed to the other compound produced by the chemical reaction of NaCl at HPHT, and the candidate is possibly the Na-rich member of the Na-Cl system.

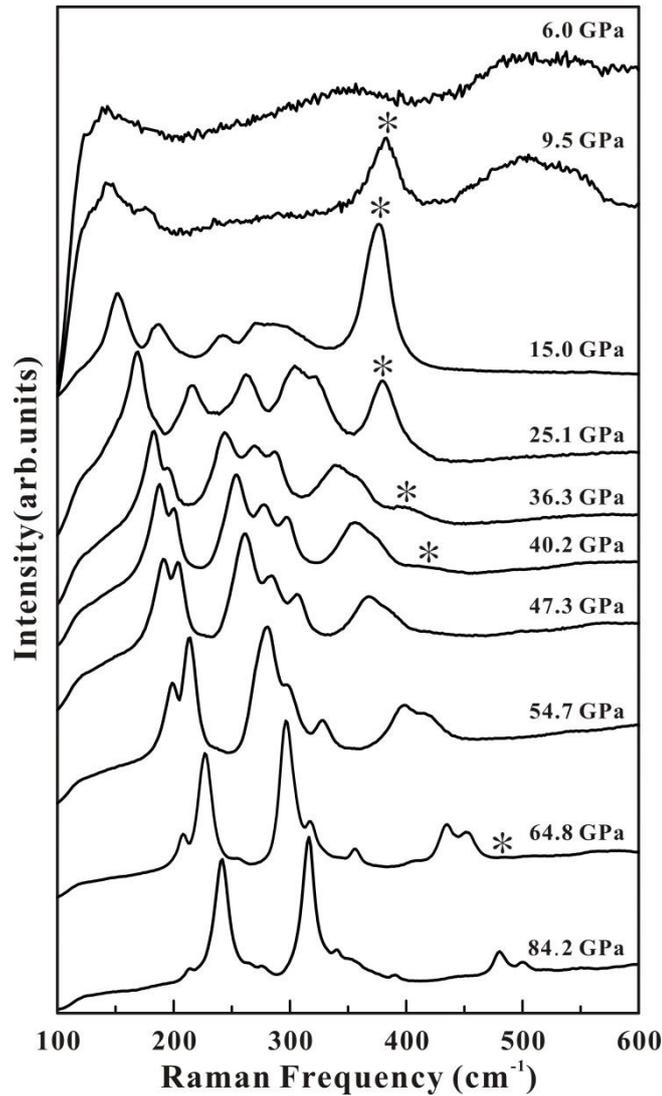

**Fig. 4.** Several representative Raman spectra measured at various pressures on decompression after annealing at about 85 GPa. It shows the change of Raman bands with pressure. The particular band marked by asterisk exhibiting obvious intensity change with pressure may be attributed to *P*4/*mmm*-Na$_3$Cl, and the other bands belong to *Pnma*-NaCl$_3$.

Although several compounds such as Na$_3$Cl, Na$_2$Cl and Na$_3$Cl$_2$ have been predicted to be stable at different pressure regions, the only synthesized Na-excess member in the Na-Cl system is the tetragonal *P*4/*mmm*-Na$_3$Cl. It is prepared by the reaction of NaCl + Na at HPHT [15]. To verify the Na-rich member of the Na-Cl system produced in this study, we present several Raman spectra measured at various pressures (curve a and b) comparing with the observed 40 GPa Raman spectrum referred to as that of *P*4/*mmm*-Na$_3$Cl (curve c) shown in supplementary materials Fig. S11 in reference 15, as shown in Fig. 5. It is surprising that both spectra of b and c show the nearly identical profile, even though they are measured in different sample systems, i.e., one is only NaCl and the other NaCl + Na. It suggests that the Raman bands of spectrum c, except the band marked by asterisk, are attributed to *Pnma*-NaCl$_3$; and the weak band marked by asterisk is possibly resulted from the *P*4/*mmm*-Na$_3$Cl [15]. This discussion, although different from Zhang et al [15], is supported by the group theory analysis of *P*4/*mmm*-Na$_3$Cl having two Raman active modes ($E_g + A_{1g}$). We will discuss the unexpected occurrence of *Pnma*-NaCl$_3$ in the NaCl + Na experiment in the later sections.

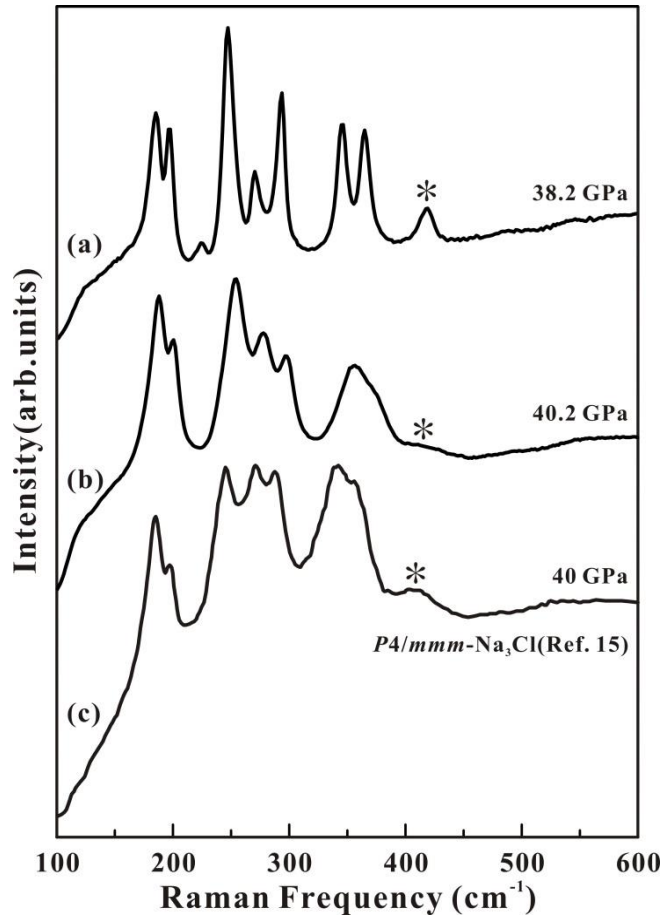

**Fig. 5.** Comparing the experimental Raman spectra (a) and (b) with (c) cited from reference 15

was measured at 40 GPa in the NaCl + Na experiment to determine the occurrence of *P*4/*mmm*-Na$_3$Cl by the decompression reaction of NaCl at HPHT. The similarity of spectrum (c) with spectrum (a) and (b) demonstrates the occurrence of *Pnma*-NaCl$_3$ in the NaCl + Na experiment. It suggests that the particular band marked by asterisk in the Figures can be referred to as the mode of *P*4/*mmm*-Na$_3$Cl.

The frequency-pressure relationship of the observed Raman bands in this study is plotted in Fig. 6. The Raman shift of each bands either on decompression (black dots) or immediately after heating (red dots) is almost identical, and the Raman bands show no obvious discontinuity as pressure changing, which indicates that no phase transition occurring for NaCl$_3$ and the possible Na$_3$Cl during decompression before they disappears.

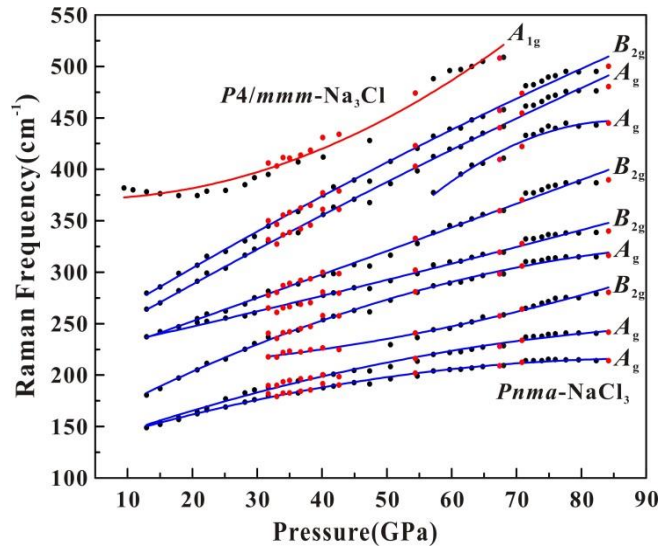

**Fig. 6.** Pressure dependence of the Raman frequencies of *Pnma*-NaCl$_3$ and *P*4/*mmm*-Na$_3$Cl. The red dots were collected just after heating, and the black dots were measured on decompression. The pressure-frequency relationship of all bands displays nonlinear feature. These data were fitted to a quadratic equation shown as blue lines for *Pnma*-NaCl$_3$ and red line for *P*4/*mmm*-Na$_3$Cl, respectively.

### *3.3. Chemical reactions of NaCl at high pressure and high temperature conditions*

Our Raman observations provide unambiguous evidences suggesting the occurrence of the NaCl$_3$ and the possible Na$_3$Cl at high pressure of 31-85 GPa after NaCl sample heated by laser to 1800±300 K. According to the Raman analysis above, we give the equation of chemical reaction of NaCl under pressure above 31 GPa and

high temperature:

4NaCl ---- NaCl$_3$ + Na$_3$Cl 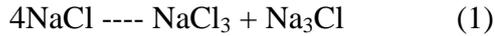 (1)

However, the convex hull diagram for Na-Cl system at various pressures [15] suggests that NaCl is the most stable compound and will not decompose spontaneously in the wide pressure region. Most of the HPHT XRD experiments by using NaCl as thermal insulator also suggested the stability of NaCl. It is distinctly inconsistent with our Raman observations. We speculate that both the non-equilibrium process in our experiments and the possible catalytic effect of the transition metal oxides of LAMs used in this study may lead to the chemical decomposition of NaCl at HPHT. The sample assembly in this study, as illustrated above, generates inevitably large temperature gradient in the NaCl sample layers as laser heating. On the one hand, the obvious temperature gradient in the NaCl layers may result in the distinctly different reactive properties and behaviors from that of in equilibrium, and therefore, the chemical decomposition of NaCl at HPHT couldn't be described by equilibrium thermodynamics. On the other hand, NaCl chemical decomposition occurs at quite high temperature as observed in this study. The large temperature gradient in NaCl sample layers may lead to a small amount of NaCl to decompose, and therefore, the reactive products may be hard to detect by XRD measurement in this kind of experiment. However, Micro-Raman spectroscopy is mainly dependent on the polarized behavior of the matter be detected, and the amount is less important relative to the polarization for the intensity of the spectrum. In addition, it is well known that the transition metal oxides generally exhibit the catalytic behavior for chemical reactions. The superposition of catalytic effect on non-equilibrium state of NaCl layers may favor to the decomposition of NaCl at HPHT.

It seems puzzling that, however, the Raman spectrum observed at 40 GPa in the NaCl + Na experiment reported by Zhang et al [15] indicates the presence of the *Pnma*-NaCl$_3$ as discussed above. Evidently it violates the theoretical prediction of the stability of the Cl-rich NaCl$_3$ in the Na-Cl system [15]. To understand the paradoxical observation, we suggest that there are two chemical reactions occurring simultaneously in the NaCl + Na experiment. The one is the normal chemical reaction

predicted theoretically and confirmed by the XRD measurement [15]:

$$NaCl + 2Na \longrightarrow Na_3Cl \quad (2)$$

and the other is the decomposition reaction of NaCl of chemical equation (1), which is also benefitted from the non-equilibrium process and the possible catalytic effect of Na metal. The feature of strong Raman bands of *Pnma*-NaCl$_3$ makes it to be observed although it may be a very small amount relative to the *P*4/*mmm*-Na$_3$Cl.

## 4. conclusions

The Raman observations demonstrate clearly the chemical reaction of NaCl at high pressure above 31 GPa and high temperature about 1800±300 K in the laser heating DAC experiments. The Raman analysis suggests that the decomposition products of NaCl include NaCl$_3$ with orthorhombic *Pnma* structure and possible Na$_3$Cl with tetragonal *P*4/*mmm* structure. It reveals that the competition of the unique Cl-Cl covalent bond and Na-Na metallic bond with the Na-Cl ionic bond may prefer the occurrence of the NaCl$_3$ and Na$_3$Cl combination in the experimental environments. The experimental observations in this study are obviously different from the theoretical prediction about the stability of NaCl at high pressure and most of the reported DAC HPHT experiments by using NaCl as thermal insulator. We consider that the chemical reactivity of NaCl is related to the large temperature gradient in the NaCl sample layers indicating the non-equilibrium state. The non-equilibrium thermodynamics complies with a series of rules much different from that of equilibrium thermodynamics, and we speculate that the large temperature gradient may prefer to enhance the chemical decomposition of NaCl. Moreover, the LAMs used in the experiments could also catalyze the decomposition of NaCl. Considering the requirement of higher temperature and the situation of the temperature gradient in the sample assemblage in the DAC experiments, the reactive products of NaCl may be a small amount and could be hard to detect by the XRD measurements due to the detected limit; however, the micro-Raman spectrum is sensitive to the polarization of materials be detected and the amount is relatively less important for the spectrum intensity.

Alkali halides usually show similar properties due to the analogous valence electronic structure. We consider that the chemical reaction could also appear in other alkali halides such as KCl and KBr at high pressure and high temperature in the non-equilibrium conditions. It is a new approach to investigate the distinctive non-1:1 stoichiometric alkali metal-halogen compounds that may display the exotic properties.

**Acknowledgments**

This work was supported by the Strategic Priority Research Program (B) of the Chinese Academy of Sciences (XDB18010403), and the National Natural Science Foundation of China (41572030).